\documentclass[%
 reprint,
 amsmath,amssymb,
 aps,
]{revtex4-1}
\usepackage{graphicx}
\usepackage{dcolumn}
\usepackage{bm}
\graphicspath{ {Figures/} }

\usepackage{subfigure}
\usepackage{color}
\usepackage{float}
\usepackage{flafter}

\begin{document}

\preprint{APS/123-QED}

\title{Experimental investigation of  amplification, via a mechanical delay-line, in a rainbow-based metasurface for energy harvesting}

\author{Jacopo M. De Ponti$^{1,2}$, Andrea Colombi$^3$, Emanuele Riva$^{2}$, Raffaele Ardito$^{1}$, Francesco Braghin$^{2}$, Alberto Corigliano$^{1}$ and Richard V. Craster$^{4,5}$
}
\affiliation{$^1$ Dept. of Civil and Environmental Engineering, Politecnico di Milano, Piazza Leonardo da Vinci, 32, 20133 Milano, Italy}
\affiliation{$^2$ Dept. of Mechanical Engineering, Politecnico di Milano, Via Giuseppe La Masa, 1, 20156 Milano, Italy}
\affiliation{$^3$ Dept. of Civil, Environmental and Geomatic Engineering, ETH, Stefano-Franscini-Platz 5, 8093 Z\"urich, Switzerland}
\affiliation{$^4$ Department of Mathematics, Imperial College London, London SW7 2AZ, UK}
\affiliation{$^5$ Department of Mechanical Engineering, Imperial College London, London SW7 2AZ, UK}

\begin{abstract}
\noindent

We demonstrate that a rainbow-based metasurface, created by a graded array of resonant rods attached to an elastic beam, operates  as a mechanical delay-line by slowing down surface elastic waves to take advantage of wave interaction with resonance. Experiments demonstrate that the rainbow effect reduces the amplitude of the propagating wave in the host structure. At the same time it dramatically increases both the period of interaction between the waves and the resonators, and the wavefield amplitude in the rod endowed with the harvester. Increased energy is thus fed into the resonators over time, we show the  enhanced energy harvesting capabilities of this system.
\end{abstract}

\maketitle

\noindent 
%
The opportunity to realize novel metamaterial devices in mechanics has recently attracted growing interest within the research community. In the context of elastic waves, considerable effort has been devoted to the creation and design of novel wave control devices motivated by the numerous applications of technological relevance involving vibrations, such as nondestructive evaluation and acoustic isolation. Topical examples include scattering-free waveguides \cite{PhysRevX.8.031074,vila2017observation}, elastic wave diodes \cite{PhysRevApplied.13.031001}, acoustic absorption and control  \cite{PhysRevB.95.014205,PhysRevB.99.174109,PhysRevApplied.7.054006} and for waves in elastic plates a range of wave manipulation devices \cite{PhysRevApplied.8.054034,PhysRevLett.119.034301,PhysRevAppl.11.014023} have been developed. 

Metamaterials are often employed in combination to create  multi-physics materials that leverage energy conversion phenomena between mechanical deformations and, for instance, electrical stimuli via piezoelectric coupling. This strategy has been exploited in the past to obtain multifunctional devices \cite{Sugino}. In this context, among the promising functionalities of acoustic metamaterials, focusing and mirroring of energy \cite{CrasterBook,PhysRevB.99.220102} have been successfully employed for flexural waves \cite{jasa.137.1783} 
and extended to vibration energy harvesting using piezoelectric materials \cite{MirrorEH,LensEH,PhysRevApplied.10.024045,Carrara}.
 
A versatile way to manipulate elastic surface waves leverages graded metasurfaces, which are structures incorporating the gentle variation of resonating elements placed on, or within, the surface. This modulation strategy was conceived, and functionally designed, to achieve broadband and efficient performance in deep elastic substrates (half-spaces) for surface elastic Rayleigh wave trapping \cite{Colombi1} and for thin elastic plates demonstrating lensing \cite{Colombi3} 
through a phenomenon generally known in physics as the \textit{rainbow effect} \cite{Hess}. A peculiar characteristic of these systems, for elastic half-spaces, is their mode conversion capability allowing surface Rayleigh waves to hybridise into bulk waves within the  half-space. This mechanism, initially discovered for Rayleigh and shear waves \cite{Colombi1}, has been recently generalized to pressure waves \cite{Greg1} thereby  enlarging the range of possible applications in geophysics, ultrasonic inspection and surface acoustic wave (SAW) devices. These all rely upon grading the system and obtaining frequency selective spatial separation; 
initially explored in electromagnetism in order to tailor a negative-index ‘\textit{left-handed}’ waveguide \cite{Hess}, the rainbow effect has been later investigated in acoustics \cite{Zhu}, water waves \cite{Bennetts} and fluid loaded plates \cite{Skelton}, amongst others.
 
Here we experimentally explore the performance of a graded metasurface, proposed in \cite{DePonti}, which is functionally designed to enhance the energy harvesting from a host structure. The graded array of resonating rods  is employed to manipulate the wave dispersion, which  can be interpreted as a wavenumber transformation induced along the structure. This effect, on one hand, allows us to slow-down waves and \textit{de facto} increases the interaction time between the wave and the harvester, clamped on the structure. On the other hand, the wavenumber transformation is accompanied by an amplitude decrease in the waveguide together with a strong amplification inside the rods. The interplay of these two effects, namely amplification and deceleration, is important, in terms of performance, to understand  and experimentally we illustrate that, for a sufficiently long excitation time, the mechanical delay line is able to harvest six times the energy of a single harvester when they are attached to the same structure. 

\begin{figure*}[t!]
\centering
    \subfigure[]{\includegraphics[width=0.39\textwidth]{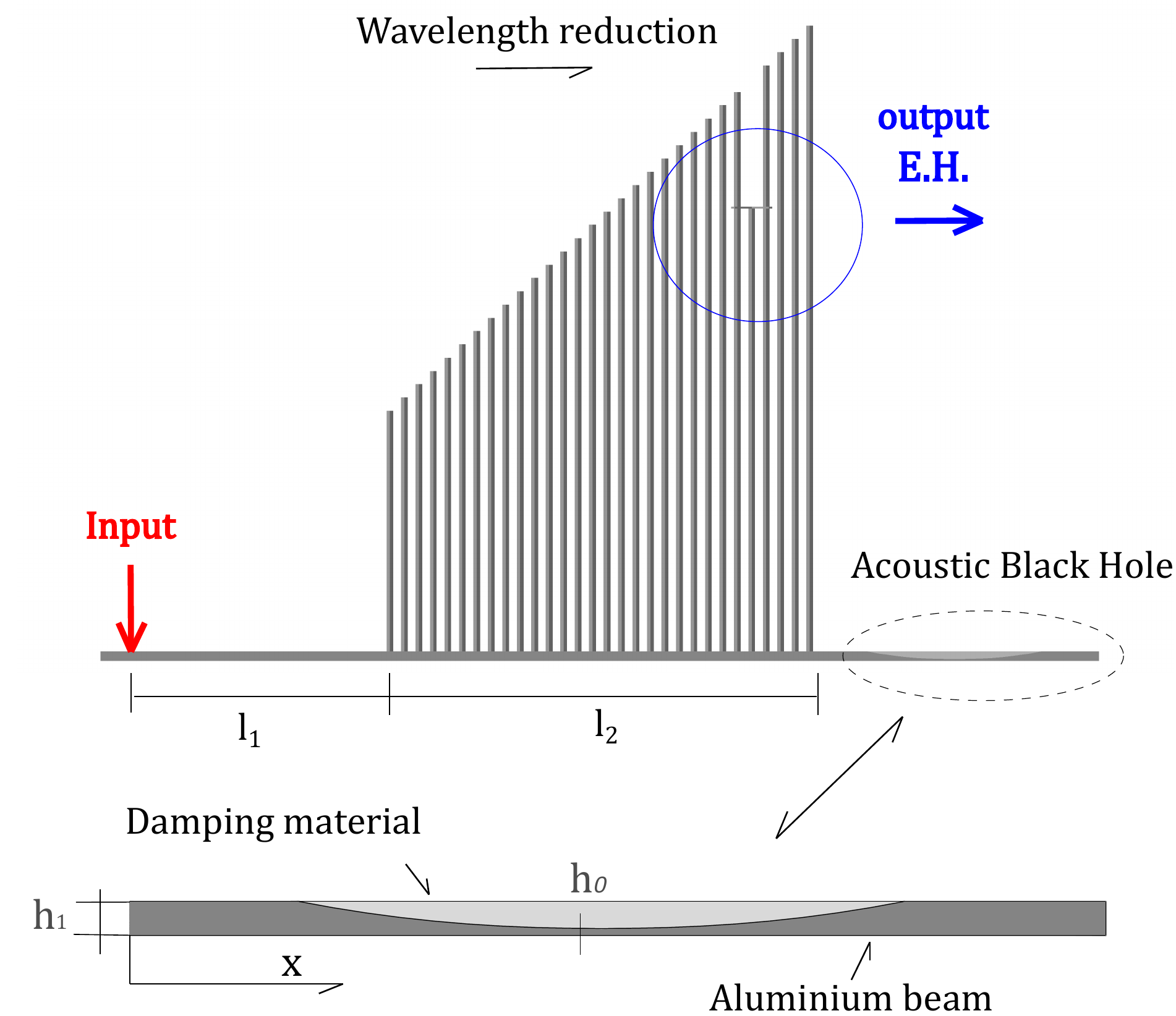}}
    \subfigure[]{\includegraphics[width=0.45\textwidth]{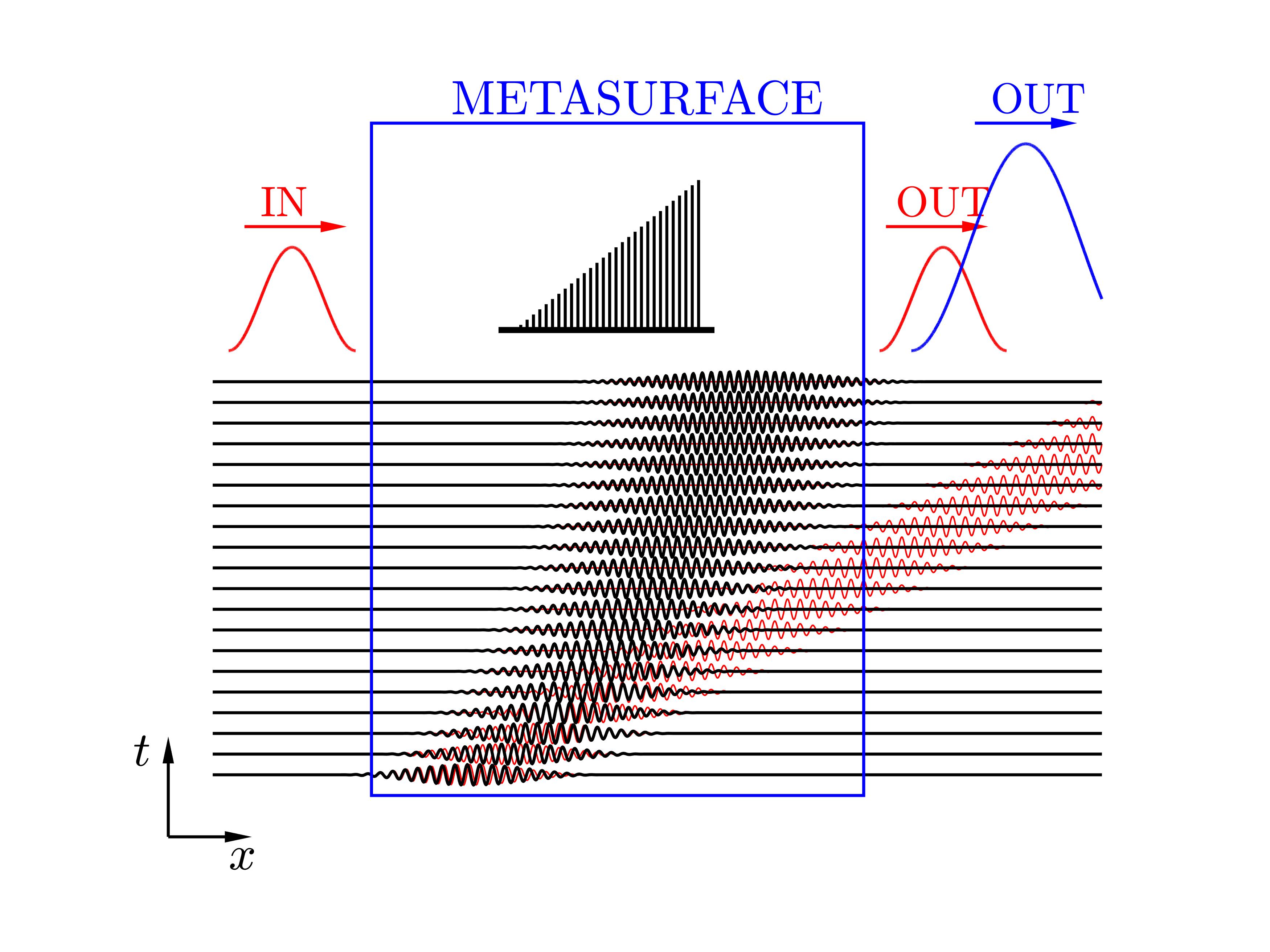}}
    \caption{(a) Schematic of the system. The rainbow device is made of an array of resonant rods, which are clamped on a host beam. The input region is represented in red. The output region, corresponding to a rod equipped with a harvester is represented in blue. In the bottom part of the figure, a detailed view of an acoustic black hole that emulates absorbing boundary conditions is shown. (b) Schematic of the desired metasurface effect on the wave propagation. The red curve is a schematic representation of the output achievable through a conventional harvester, i.e. without metasurface. The blue curve represents the amplification that can be obtained through a rainbow device.}
	\label{fig:Schematic}
\end{figure*}

\begin{figure*}[!t]
\centering
	\subfigure[]{\includegraphics[width=0.19\textwidth]{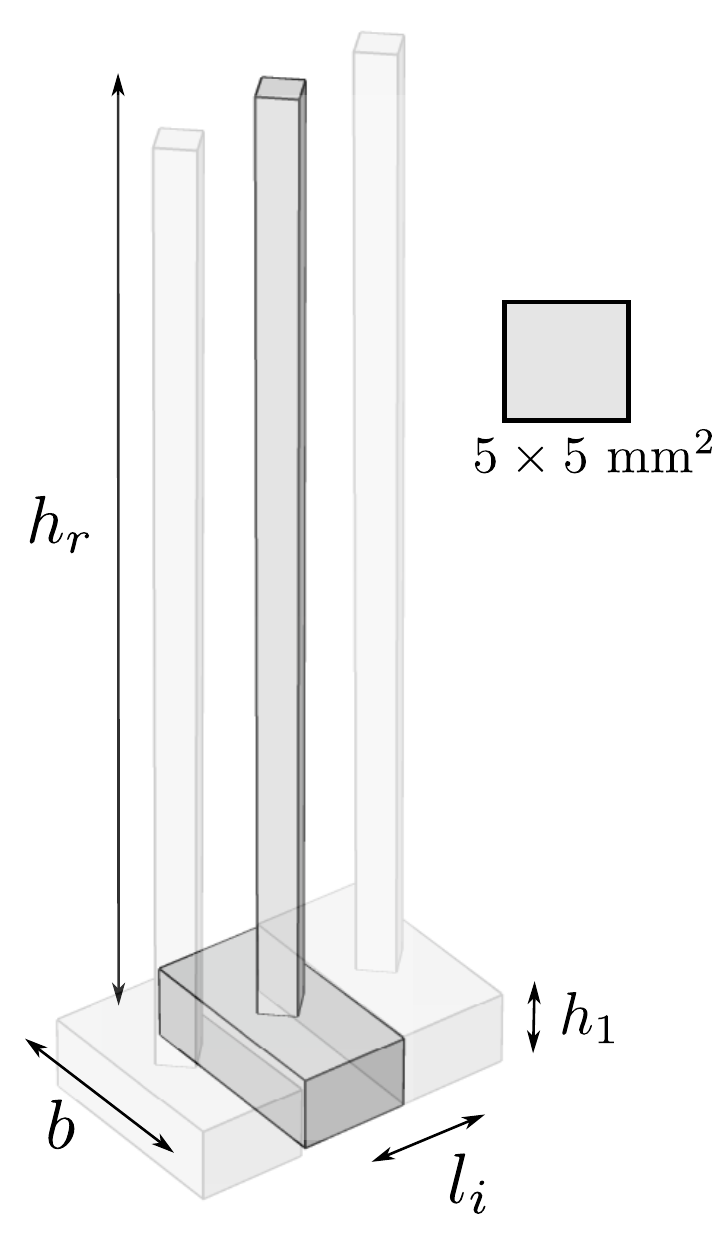}}
	\subfigure[]{\includegraphics[width=0.8\textwidth]{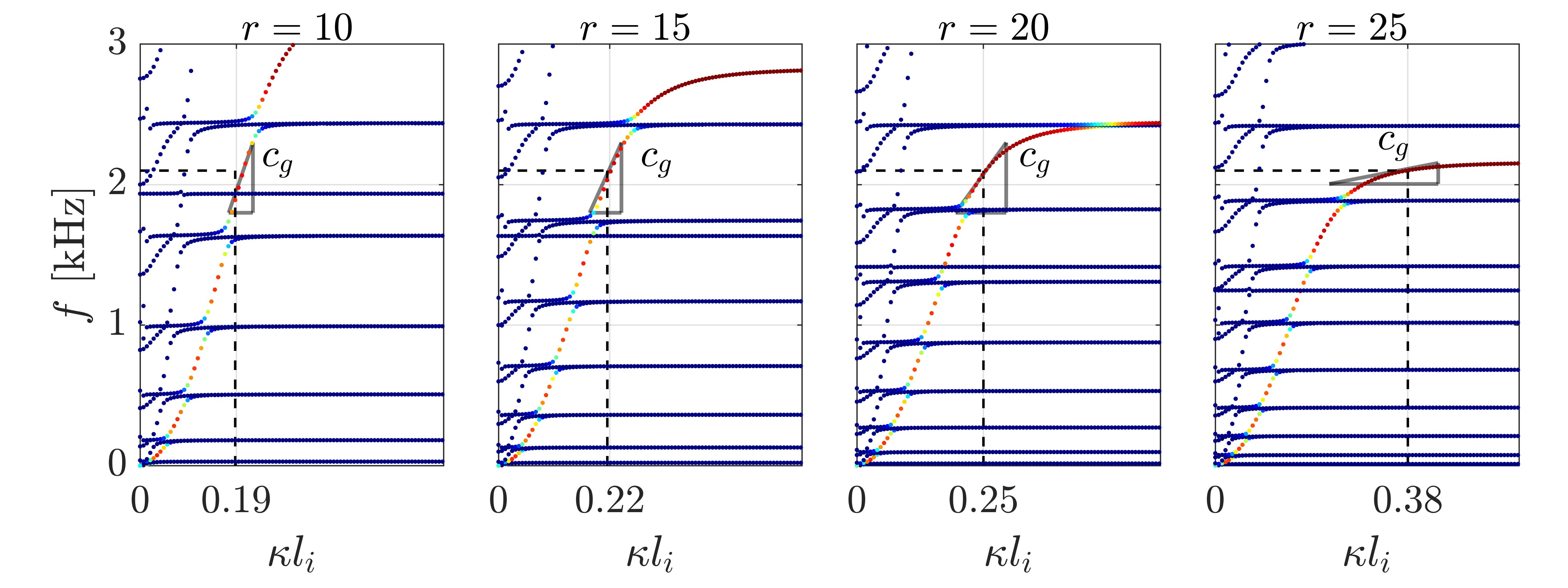}}
	\caption{(a) Schematic of the unit cell. (b) Dispersion relation associated to the unit cell made of the host beam equipped with the  $r^{th}$ resonator, obtained using COMSOL Multiphysics{\textsuperscript{\textregistered}}. The colors represent the wave polarization factor $p=|w|^2/(|w|^2+|u|^2+|v|^2)$ (where $u,v,w$ are the displacements in the $x,y,z$ directions), computed through numerical integration within the resonator spatial domain. The wavenumber and associated group velocity are highlighted in the neighborhood of the working frequency with dashed lines and solid lines, respectively.}
	\label{fig:Dispersion}
\end{figure*}

%
We consider the graded metasurface illustrated in Fig.~\ref{fig:Schematic}(a); this is constructed using an aluminum beam ($E_a=70\:GPa$, $\nu_a=0.33$ and $\rho_a=2710\:kg/m^3$), having thickness $h_1=10$ mm and width $b=30$ mm, which is augmented by an array of $N=30$ resonating rods of variable height and square cross-section area $5$ mm $\times\;5$ mm. The array of consecutive rods is separated by a constant lattice spacing of $l_i=15$ mm, which is sub-wavelength compared to the operational frequencies of the device, and the array is placed at $l_1=300$ mm from the input source. 

\begin{figure}[h!]
\centering
	\includegraphics[width=0.43\textwidth]{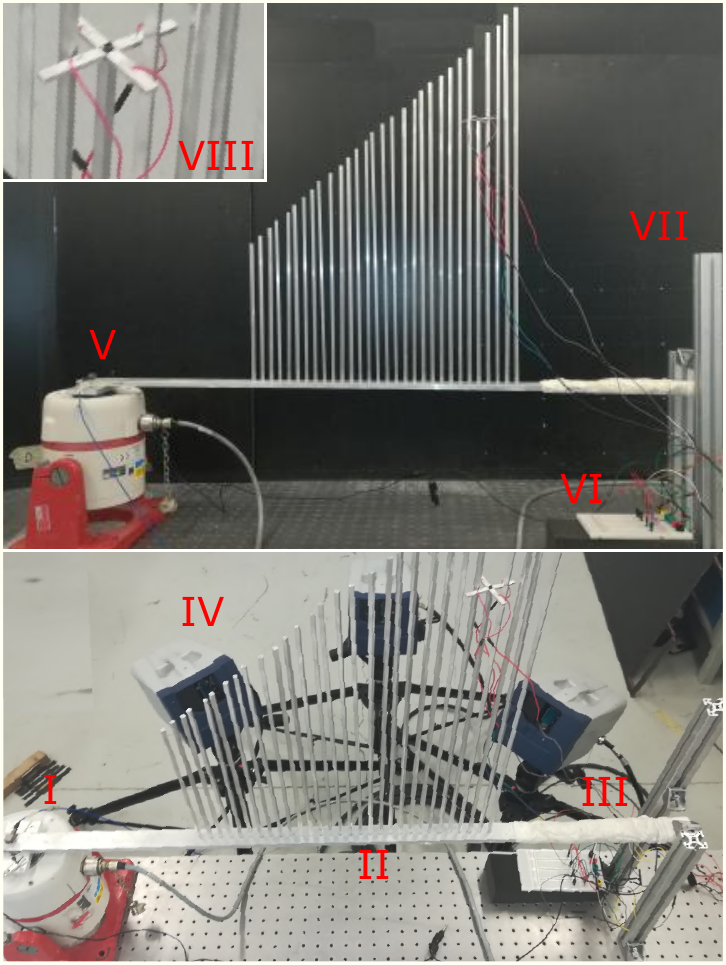}
	\caption{Experimental setup. The excitation is provided through an electrodynamic shaker (I). The rainbow metasurface is mounted on an otherwise plain beam (II) supplemented by an acoustic black hole (III), to prevent spurious reflections, and a 3D laser vibrometer used to measure the wavefield on the bottom surface of the beam (IV). The input acceleration is measured through an accelerometer located alongside	the shaker (V) while the output power is measured by connecting the piezo-electrodes to a passive resistive load. The system is suspended through elastic cables located alongside the acoustic black hole (VII). A zoomed-in view of the harvester is shown in (VIII).}
	\label{fig:Setup}
\end{figure}

The dynamical behavior of the overall system is controlled by tuning the longitudinal resonance of each rod, to functionally provide filtering behavior for wave propagation and modified dispersion in the neighborhood of the resonance frequency. Here, the operational frequency of each resonator is smoothly varied along the beam length, this slows down the wave propagation through the array and increases the interaction time between the wave and the structure, as shown in the schematic of Fig. \ref{fig:Schematic}(b); for the host beam, wave propagation can only be modified by the dispersive nature of the medium (red curve), in contrast a carefully designed metasurface is able to dramatically decrease the speed of the wave (black curve). This behavior is critical in achieving efficient electromechanical energy conversion between an input source and a piezoelectric harvester (red and blue arrows in Fig. \ref{fig:Schematic}(a), respectively) mounted on one of the resonators.

Dispersion curves are an important tool in terms of understanding the mechanism and designing the array and they are shown in Fig. \ref{fig:Dispersion}, in which a unit cell dispersion relation is investigated for different rod heights $h_r=h_i+r(h_f-h_i)/N$ spanning the range between $h_i=250$ mm and $h_f=650$ mm. These dispersion curves are for an infinite array of identical rods, each of the same height and equally spaced, and this local dispersion behaviour allows us to understand how the graded array operates at the local spatial position corresponding to each height; implicitly the assumption is that nearest neighbours dominate the behaviour of each rod and that the grading is gradual. The varying rod heights, $h_r$, each yield different dispersion curves and, therefore, induces a different local wave speed, which is hallmark of the rainbow effect. For an operating frequency in the neighborhood of $f=2.05$ kHz, the modulation is therefore accompanied by a wavenumber transformation (highlighted with dashed lines in Fig. \ref{fig:Dispersion} when the smooth modulation of the rod profiles are considered. i.e. when the parameter space $\left[h_i,h_f\right]$ is mapped within the physical space $x\in\left[l_1,l_1+l_2\right]$.
Consistent with the description using dispersion, the group velocity $c_g=\partial\omega/\partial\kappa$ is modified by the changing rod heights and it is illustrated in Fig. \ref{fig:Dispersion} through solid lines. Specifically, $c_g$ decreases for increasing height of the rods $h_r$ until it is nullified for the $27^{th}$ resonator, which is a key observation and hereafter employed to enhance the harvested power of the $26^{th}$ resonator. The final three resonators in the array have heights such that, at the operating frequency, we are in the band-gap and wave propagation is disallowed and this acts to prevent wave propagation away from the harvester. To be precise, the mechanism we have identified is rainbow reflection, as theoretically explained in ref. \cite{Greg2}, since the zero group velocity and the strong impedance contrast is achieved at the rod's resonant frequency, namely at the bandgap opening \cite{Martin}.

For the experimental setup, shown in Fig. \ref{fig:Setup},  we have an array of resonators clamped on the host beam through a set of screws and spaced by $l_1=300$ mm from the left boundary.
 At the left boundary a \textit{LDS v406} electrodynamic shaker is rigidly connected to the beam, to provide excitation, and an accelerometer, \textit{PCB 352C33}, is placed on the surface in order to measure the real acceleration induced by the shaker.
The right boundary is endowed with an acoustic black hole to prevent any spurious reflections; it is achieved by gradually varying the beam thickness and through the addition of a dissipative material to emulate absorbing boundaries \cite{Krylov1,Krylov2,Rajagopal}.

A piezoelectric PZT-5H harvester ($E_p=61\:GPa$, $\nu_p=0.31$, $\rho_p=7800\:kg/m^3$, dielectric constant $\epsilon_{33}^T/\epsilon_0=3500$ and piezoelectric coefficient $e_{31}=-9.2\:C/m^2$) is placed at the position of the $26^{th}$ resonator, and allows to quantify the vibrational energy that is transduced during the experimental tests. Specifically, each piezoelectric electrode is attached to an optimal resistive load $R=6.8$ k$\Omega$, whereby the resulting voltage drop is acquired in real time. A detailed description of the black hole geometry and EH details are provided in the supplementary material (SM)~\cite{SM}.

The wave propagation is measured through a Polytec 3D \textit{Scanner Laser Doppler Vibrometer} (SLDV), which is able to separate the out of plane velocity field in both space and time. A narrow-band spectrum excitation of central frequency at $2.05$ kHz and width  $\Delta f=0.07$ kHz is synchronously started with the acquisition which, in turn, is averaged in time to decrease the noise. Two different testing conditions are considered: \textit{(a)} only one resonator, equipped with the harvester, is placed on the beam; \textit{(b)} all the resonators are present, whereby the metasurface enables the wave speed control.  This enables us to quantify the difference between a lone single harvester and the harvester embedded within the metasurface. 

\begin{figure*}[ht!]
\centering
    \subfigure[]{\includegraphics[width=0.44\textwidth]{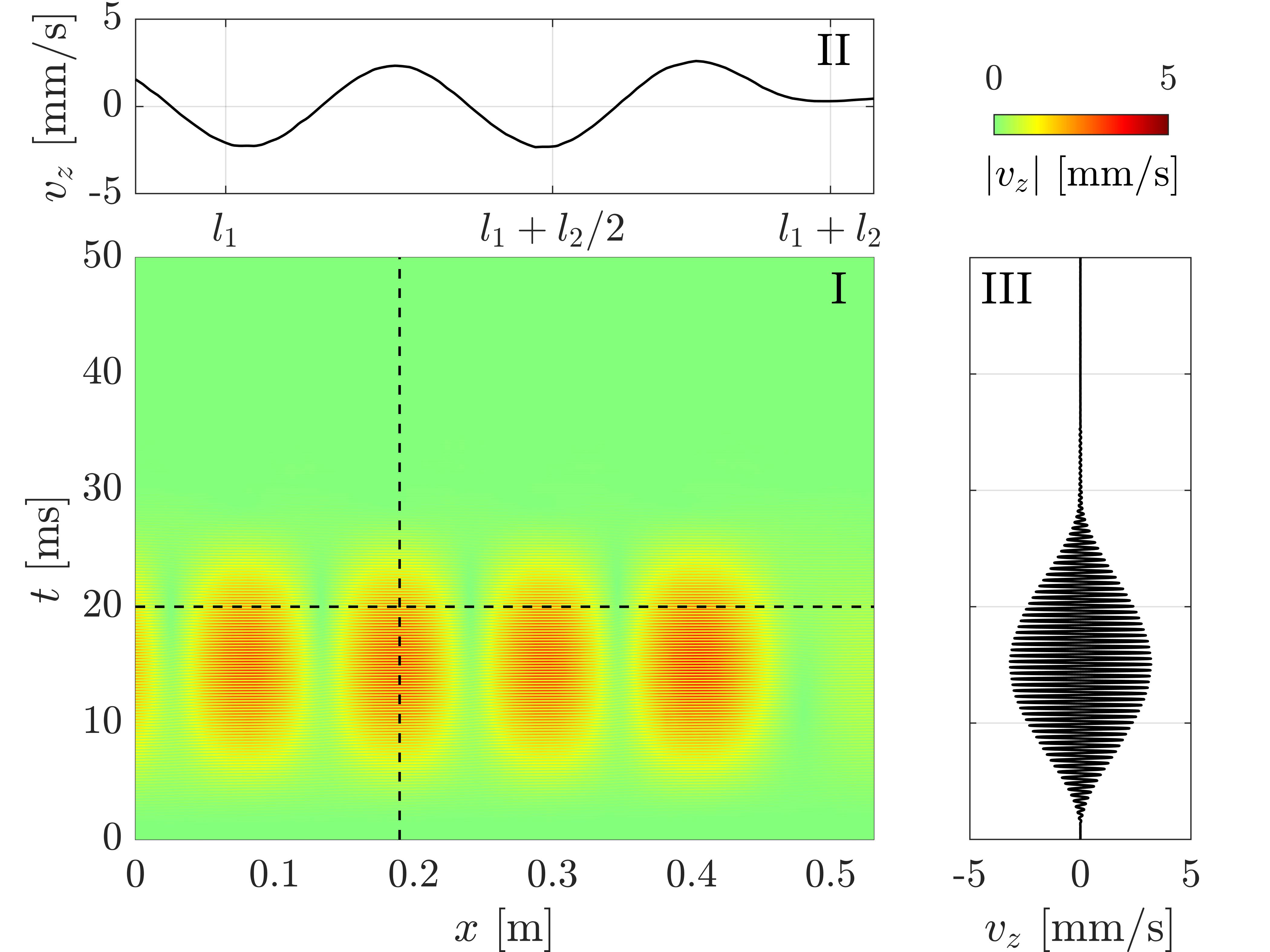}}
    \subfigure[]{\includegraphics[width=0.44\textwidth]{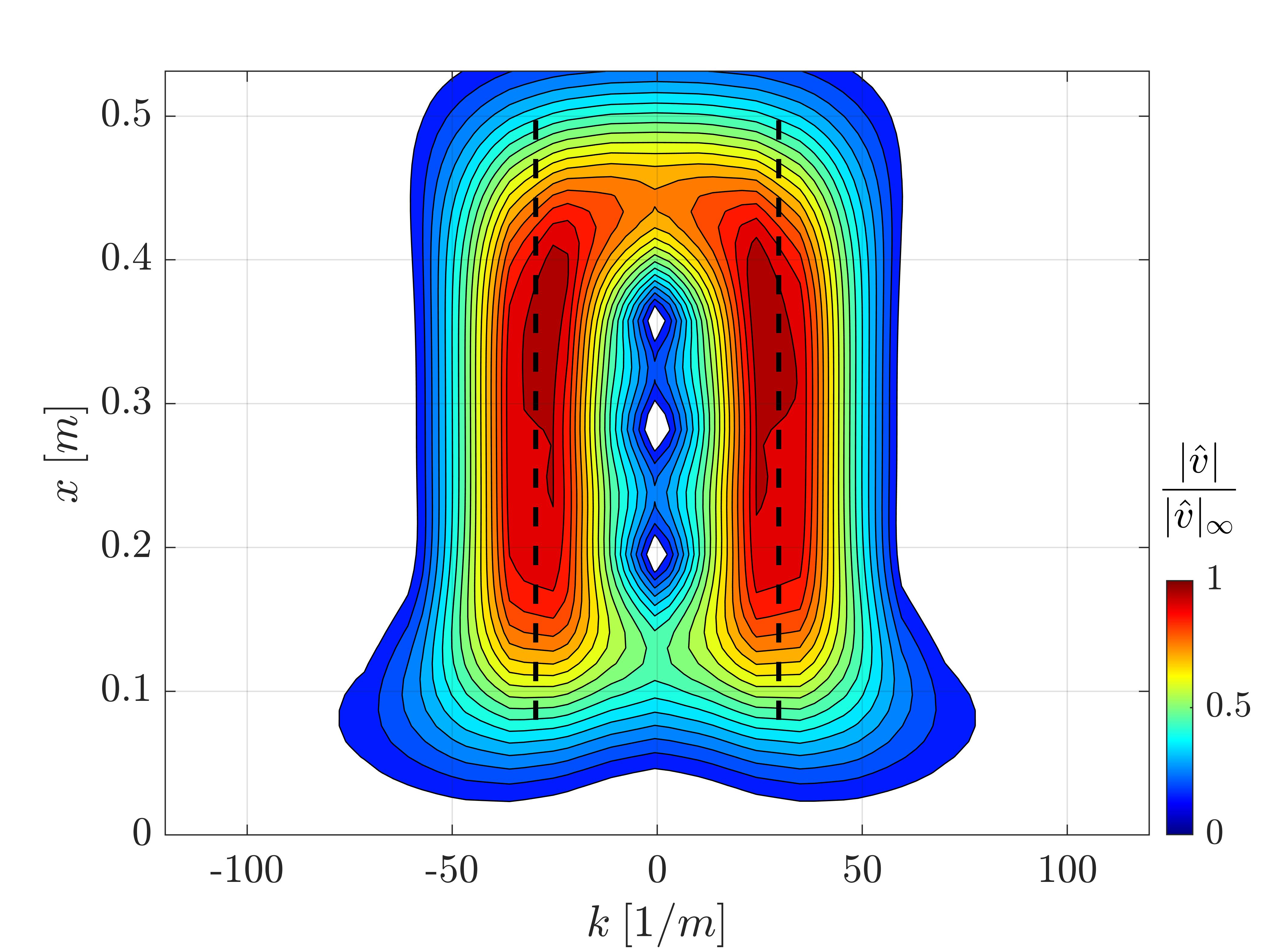}}\\
    \subfigure[]{\includegraphics[width=0.44\textwidth]{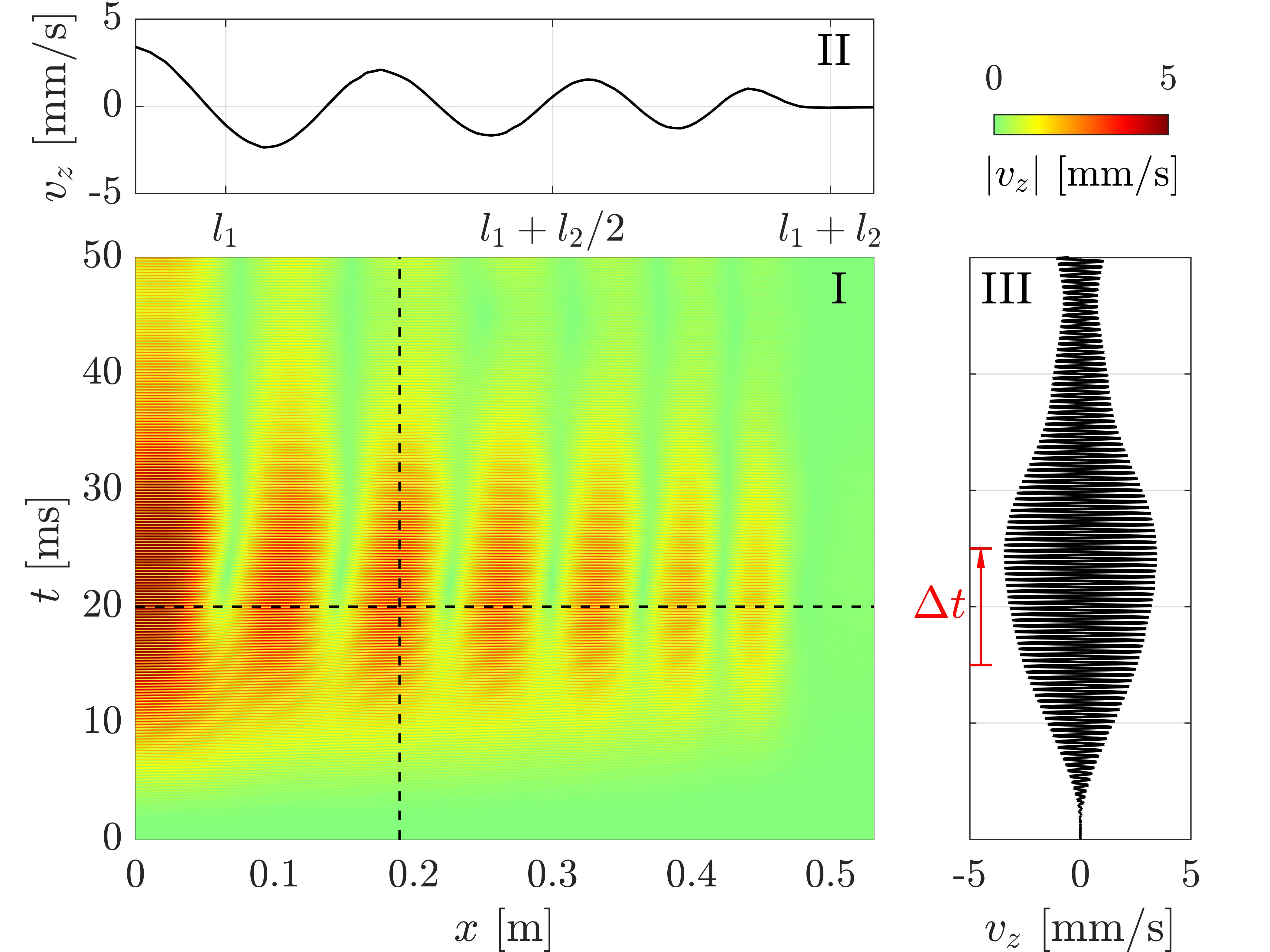}}
    \subfigure[]{\includegraphics[width=0.44\textwidth]{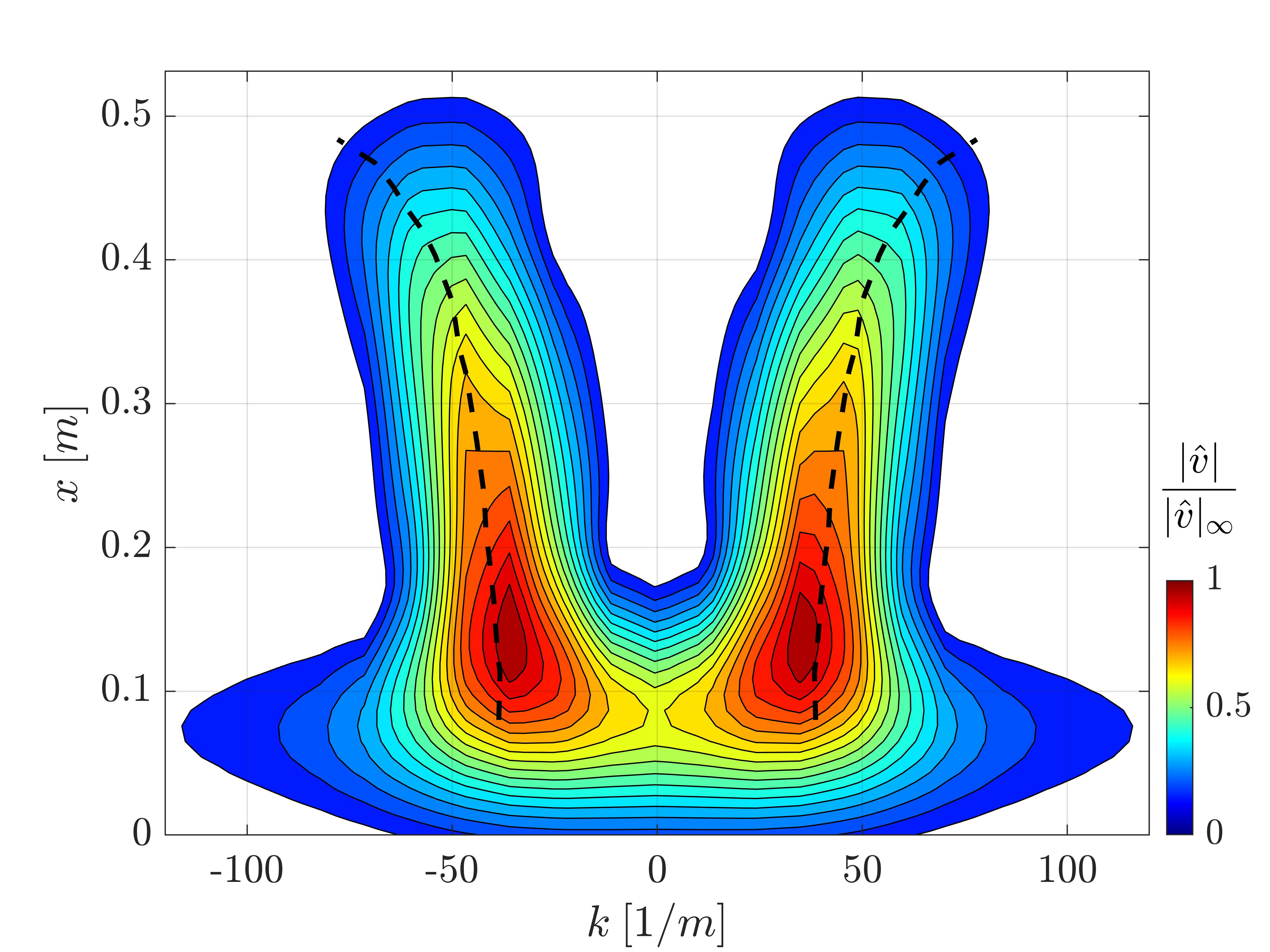}}
    \caption{Experimental velocity field for a beam with (a) one resonating rod and (c) the metasurface. The horizontal dashed lines are representative of the time instant $t=20$ ms, corresponding to the wave profile displayed on the top of the figure (${\rm II}$). The vertical dashed line is representative of the temporal wave profile measured at $x=0.2$ m and illustrated in the right side of the figure (${\rm III}$). Corresponding spectrograms for (b) single resonator and (d) the metasurface, with superimposed numerical dispersion for constant frequency and different resonator height (dashed black line). While for the single harvester the wavenumber is constant in space, a relevant wavenumber transformation can be observed when the metasurface is mounted on the beam.
    }
	\label{fig:wavefield}
\end{figure*}

\begin{figure*}[t!]
\centering
    \subfigure[]{\includegraphics[width=0.44\textwidth]{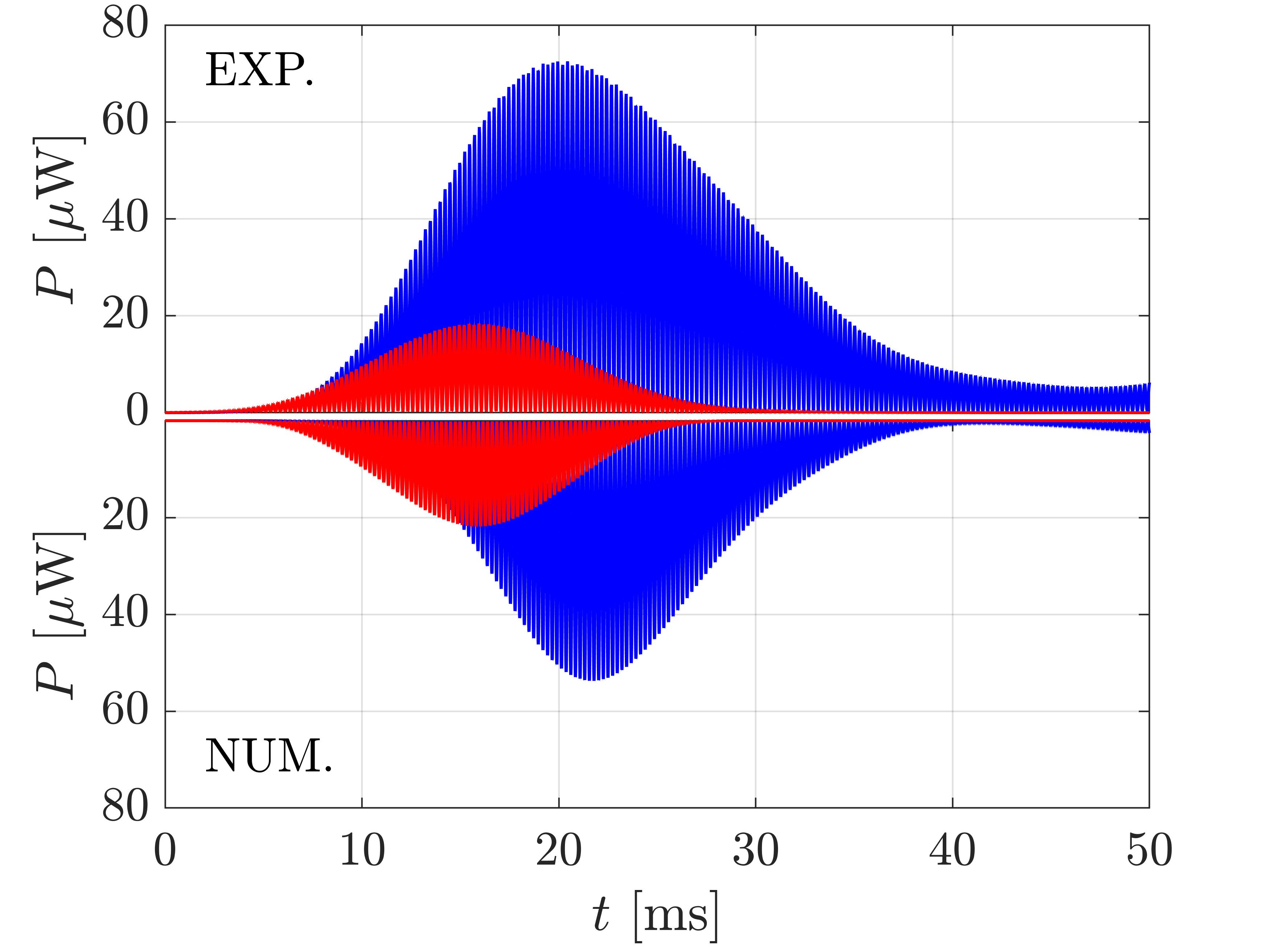}}
    \subfigure[]{\includegraphics[width=0.46\textwidth]{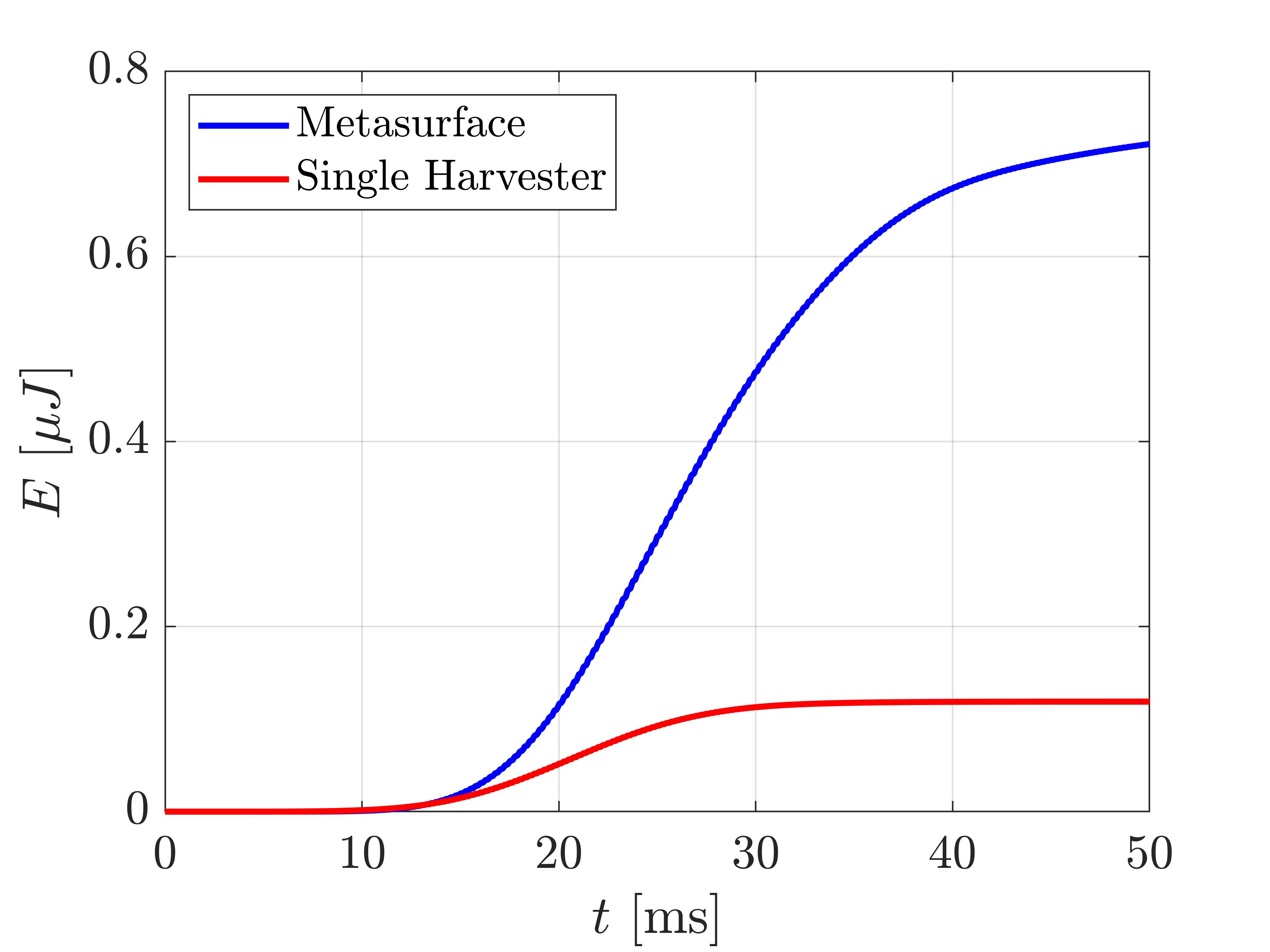}}
    \caption{(a) Numerical and experimental electric power produced by the harvester with (blue) and without (red) the metasurface. (b) Experimental cumulated electric energy in time.}
	\label{fig:Numerics}
\end{figure*}

The resulting velocity field, for the single rod, is illustrated in Fig. \ref{fig:wavefield}(a)-{$\rm I$}; since the beam is characterized by constant material properties along its main dimension, the associated wavelength is invariant in space, as is clearly visible in the velocity profile in Fig. \ref{fig:wavefield}-{\rm II}, corresponding to the time instant illustrated with the horizontal dashed line. In addition, at a generic point belonging to the beam, the imposed wave drops to zero after approximately $25$ ms, which is  highlighted in the vertical cut of the velocity field shown in Fig. \ref{fig:wavefield}-{\rm III}.

The corresponding spectrogram is shown in Fig. \ref{fig:wavefield}(b); it is computed through a 2D \textit{Fourier Transform} (FT) of the velocity field, properly windowed with a moving Gaussian function along x, which results in the function $\hat{v}\left(\kappa,x,f\right)$. The dependence upon frequency is eliminated by taking the RMS value in time, which allows to define the spectrogram as the amplitude $|\hat{v}\left(\kappa,x\right)|$, displayed with colored contours in Fig. \ref{fig:wavefield}(b),  confirming that wave propagation occurs without wavenumber transformation \cite{Riva}. 
Turning to the harvester embedded within the metasurface,  the effect of the metasurface is visible in Fig. \ref{fig:wavefield}(c)-{$\rm I$} and in the corresponding horizontal (Fig. \ref{fig:wavefield}(c)-{$\rm II$}) and vertical (Fig. \ref{fig:wavefield}(c)-{$\rm III$}) cuts. Interestingly, the associated wavelength varies along the $x$-direction, whereby its variation is accompanied by amplitude decrease, as expected from the arguments presented earlier. 
We also observe that the temporal response is delayed and that the energy remains in the system for a longer time compared to the single harvester, which further enhances the metasurface effect. This effect is created by the partial wave scattering and wave confinement occurring inside the array, and this  is further  confirmed by comparison between the expected wavenumber transformation (obtained with the numerical model and represented with black dashed lines) and the experimental spectrogram in Fig. \ref{fig:wavefield}(d).

Finally, we quantify the EH capabilities of the system by comparing the experimental power output for the host beam with one resonator (red line) and the power achieved through the metasurface effect (blue line), as shown in Fig \ref{fig:Numerics}(a) (top), for the same excitation level. It is demonstrated that, thanks to the careful design of the array of rods, the wave propagation is slowed down. As a result, a greater interaction time between the wave and the harvester and a focusing generating a field amplification on the rod endowed with the harvester is guaranteed, which reflects into a greater and delayed power output.

In addition to these experimental results, a comparison with the power calculated through a numerical simulation is shown in Fig. \ref{fig:Numerics}(a) (bottom). 
The system is numerically modeled through ABAQUS CAE 2018\textsuperscript{\textregistered}, with a user subroutine introduced to include the electric resistance \cite{Gafforelli1,Gafforelli2}, and imposing the experimental acceleration time history. The output voltage is displayed in the bottom part of Fig. \ref{fig:Numerics}, for the case of one resonator (red) and of a full metasurface (blue).
The agreement between the experimental and the numerical data confirms  that the rainbow effect is the observed mechanism and that the system is operating as we predict. 

The accumulated energy for the two scenarios, the lone harvester and the harvester embedded in the metasurface, are then compared in Fig. \ref{fig:Numerics}(b) in the range $0-50$ ms, which corresponds to a total energy of $0.12$ $\mu$J and $0.72$ $\mu$J, demonstrating a strong increase of the harvesting capabilities.
The advantage of the system is evident only after a certain amount of time, and benefits from a sufficiently continuous, or long duration, source of excitation; if the excitation is discontinuous or very short, the metasurface is less efficient \cite{DePonti}.

In conclusion, we have experimentally demonstrated potential  advantages in using a rainbow-based metasurface for wave manipulation and energy harvesting. 
The metasurface capability of slowing down waves enables a longer excitation of the resonators, which is then reflected in a higher electric power production. This behavior can be suitably employed for applications involving energy harvesting and can be scaled at the micro-scale for the implementation of next generation vibration energy harvesting devices. 

\section*{Acknowledgements}

J.M.D.P thanks Politecnico di Milano for the scholarship on `Smart Materials and Metamaterials for industry 4.0' and G. Borghi for the prototype realization. \\
The Italian Ministry of Education, University and Research is acknowledged for the support provided through the Project ``Department of Excellence LIS4.0 - Lightweight and Smart Structures for Industry 4.0”. 
R.V.C is funded by the UK Engineering and Physical Sciences Research Council (EP/T002654/1) and also acknowledges funding from the ERC H2020 FETOpen project BOHEME.

\end{document}


\preprint{APS/123-QED}

\title{Experimental investigation of  amplification, via a mechanical delay-line, in a rainbow-based metasurface for energy harvesting:
 Supplementary Material}

\author{Jacopo M. De Ponti$^{1,2}$, Andrea Colombi$^3$, Emanuele Riva$^{2}$, Raffaele Ardito$^{1}$, Francesco Braghin$^{2}$, Alberto Corigliano$^{1}$ and Richard V. Craster$^{4,5}$
}
\affiliation{$^1$ Dept. of Civil and Environmental Engineering, Politecnico di Milano, Piazza Leonardo da Vinci, 32, 20133 Milano, Italy}
\affiliation{$^2$ Dept. of Mechanical Engineering, Politecnico di Milano, Via Giuseppe La Masa, 1, 20156 Milano, Italy}
\affiliation{$^3$ Dept. of Civil, Environmental and Geomatic Engineering, ETH, Stefano-Franscini-Platz 5, 8093 Z\"urich, Switzerland}
\affiliation{$^4$ Department of Mathematics, Imperial College London, London SW7 2AZ, UK}
\affiliation{$^5$ Department of Mechanical Engineering, Imperial College London, London SW7 2AZ, UK}

\maketitle
\beginsupplement

To aid  insight  into the underlying physics of the rainbow-based metasurface, described in the main text, we give a detailed description of the experimental setup and additional numerical results.

\section{SUPPLEMENTARY NOTE 1 - FURTHER EXPERIMENTAL DETAILS}
\begin{figure*}[h!]
	\centering
	\hspace{-0.4cm}
	\subfigure[]{\includegraphics[width=0.34\textwidth] {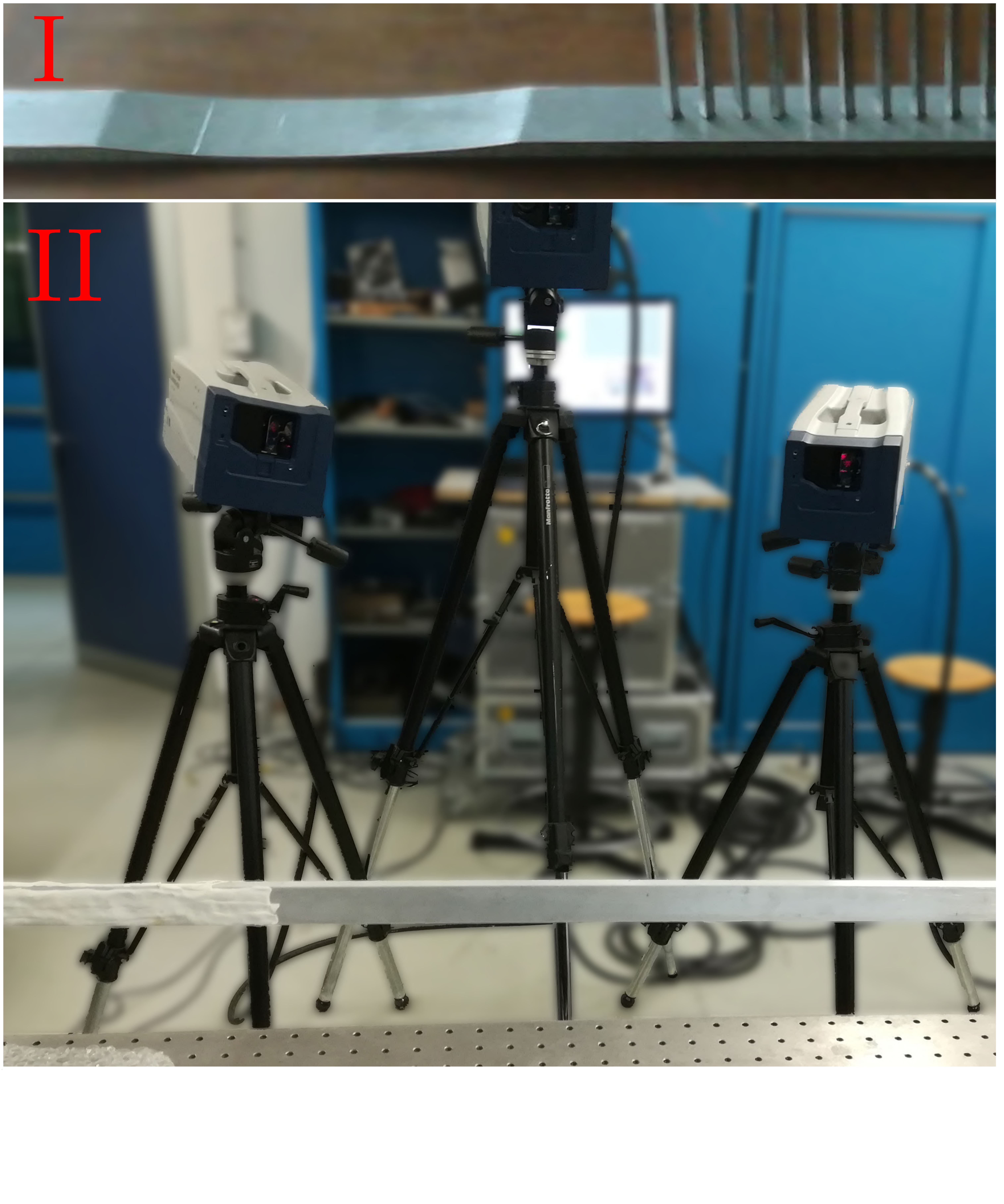} \label{FigS1a}}\hspace{1cm}
	\subfigure[]{\includegraphics[width=0.47\textwidth] {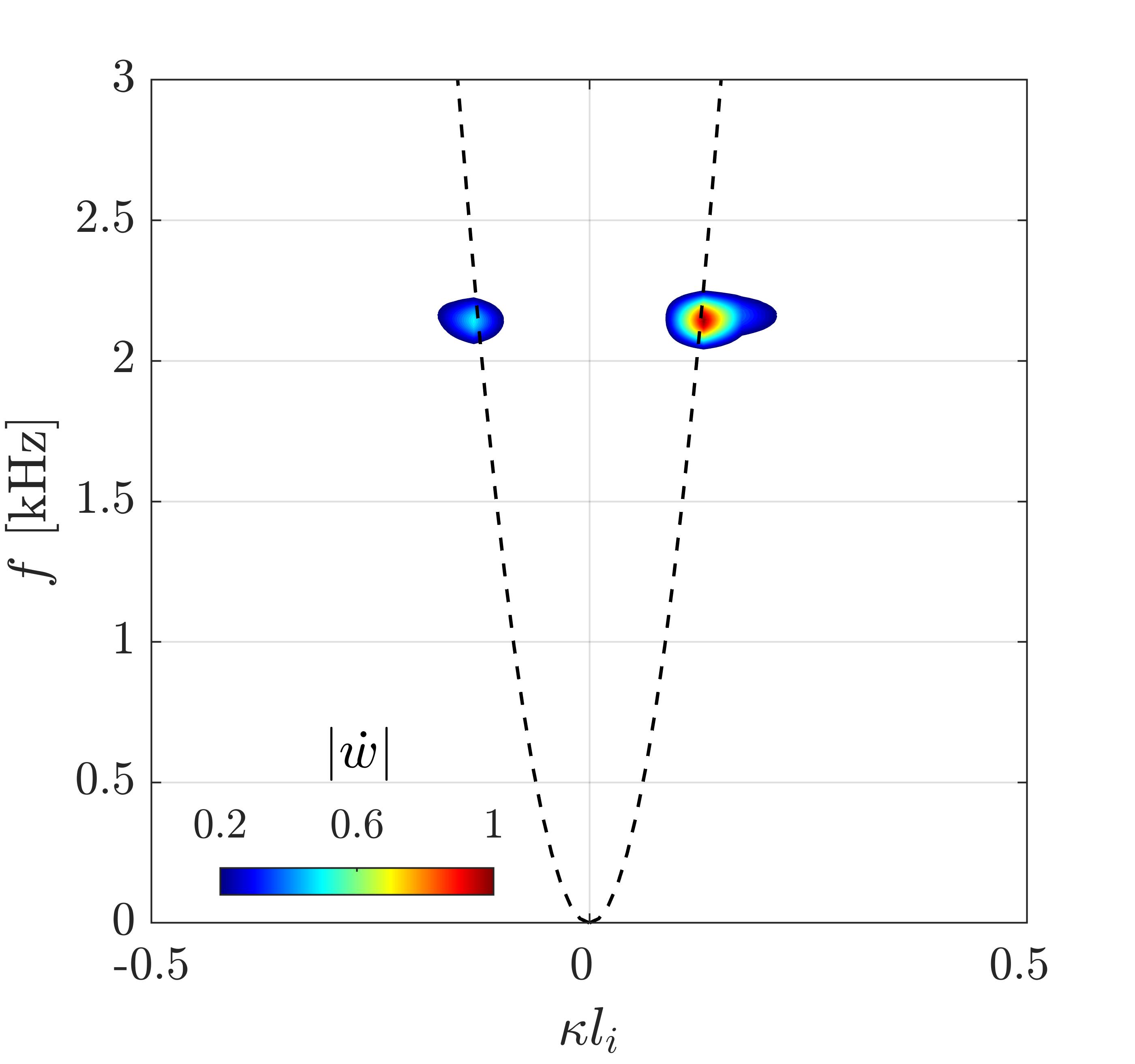} \label{FigS1a}}	\caption{(a) Experimental setup for the characterization of the acoustic black hole. The beam is clamped on the left side and the wavefield is measured on the bottom surface using a 3D scanner laser Doppler vibrometer (SLDV). The acoustic black hole is manufactured by gradually reducing the beam thickness of an otherwise constant-thickness beam and through the addition of a highly dissipative material in correspondence of the reduced cross section (I). (b) Experimental dispersion curve illustrating the input and reflected $A_0$ Lamb modes.}
	\label{Fig:ABH}
\end{figure*}

The electro-mechanical system is made of a beam augmented with $30$ resonators, which are mounted on the beam through a set of screws, emulating rigid connections. We assume that the tightening of the screws is sufficiently strong to guarantee a linear and frequency-independent behavior of the joint. The beam is also mechanically joined to a \textit{LDS v406} electrodynamic shaker at the left boundary, to provide an out-of-plane input excitation; the beam is placed perpendicularly to the excitation. 
An accelerometer, \textit{PCB 352C33}, is bonded to the beam to measure the acceleration level provided by the shaker and data is acquired by way of a National Instruments (NI) input module with internal signal conditioning.\\
The opposite end is attached to an external frame via a set of ropes, to avoid undesired deformations of the beam;  the presence of the ropes is assumed not to affect the system dynamics. 
In addition, an acoustic black hole, whose characteristics are discussed in detail below, is manufactured at the right-hand boundary to minimize any reflection of waves and to emulate absorbing boundaries.  
The $26^{th}$ resonator, that is placed at the coordinate at which the wave speed at $2.05$ kHz is minimum, is equipped with a harvester, able to convert the mechanical vibrations into electric energy. The geometrical and physical parameters of the harvester are illustrated in the final part of this section.\\ 
A Polytec 3D \textit{scanner laser Doppler vibrometer} (SLDV) measures the velocity field along the beam's main dimension, separating out the out-of-plane component. At the same time, the voltage across the harvester's electrodes is acquired through the external Polytec acquisition system. Finally, the excitation signal (provided through a \textit{KEYSIGHT 33500B} waveform generator), synchronously starts with the measurement system and consists in a input function $V(t)=V_0w(n)\sin(2\pi f_c t)$ with  amplitude  $V_0=5$ V, $w(n)$ is a Hann window, central frequency $f_c=2.05$ kHz and time duration $30$ ms; this results in a spectral content of width $\Delta_f=0.07$ kHz.
\\
\\
\\
%
\noindent
\textit{Black hole characteristics}\\
\\
To emulate absorbing boundaries we consider an acoustic black hole similar to Ref. \cite{Krylov1,Krylov2,Krylov3} and illustrated in Fig. \ref{Fig:ABH}(a)-I. Specifically, the beam cross-section is machined to achieve a variable profile for the thickness characterized by the following expression:

\begin{equation}
h(x)=\epsilon x^m + h_0,
\end{equation}

where $m=2.2$, $h_0=2$ mm and $\epsilon=1.6$. The addition of a highly dissipative material (\textit{Blu-Tack}) to the variable section provides localized damping during the wave propagation, that allows to avoid undesired reflections at the beam's free edge, therefore emulating absorbing boundary conditions. \\ 
We experimentally verified the behavior of the acoustic black hole through the setup illustrated in Fig. \ref{Fig:ABH}(a)-II, in which only the host beam is mounted on the optical table. 
The aforementioned input forcing signal ($f=2.05$ kHz, $\Delta f=0.07$ kHz) is imposed at the boundary opposite to the black hole. The experimental dispersion relation $|\dot{w}\left(\kappa,f\right)|$ is computed through the \textit{Fourier Transform} (FT) of the velocity field and compared to the analytical dispersion for the $A_0$ Lamb mode in Fig. \ref{Fig:ABH}(b). As expected, the energy content is mainly located on the positive wavenumber domain, whereas only a small amount of energy (almost $50\:\%$ of reflected component) is present in the left side.

\begin{figure*}[h!]
	\centering
	\hspace{-0.4cm}
	\subfigure[]{\includegraphics[width=0.49\textwidth] {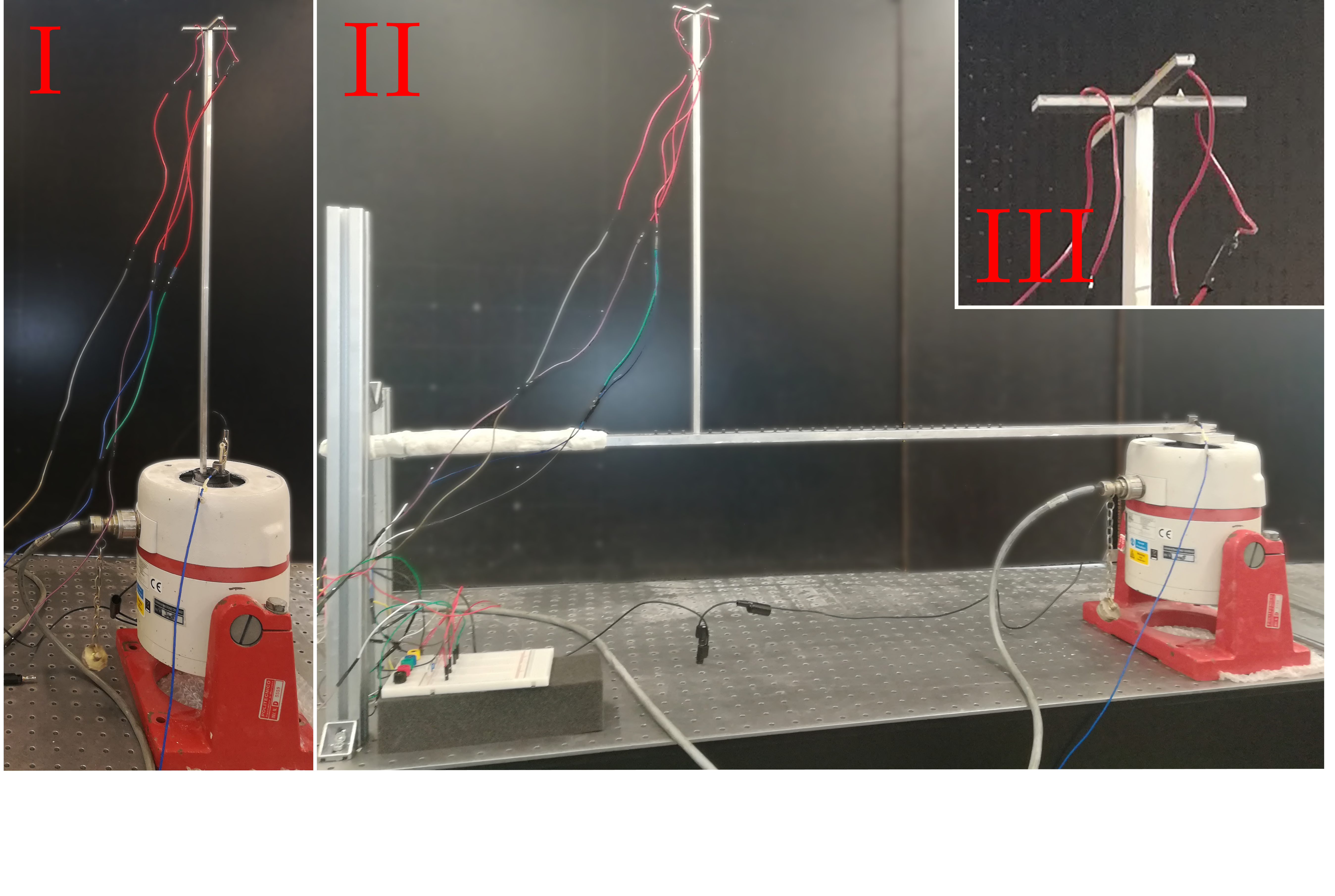} \label{FigS1a}}\hspace{0.2cm}
	\subfigure[]{\includegraphics[width=0.47\textwidth] {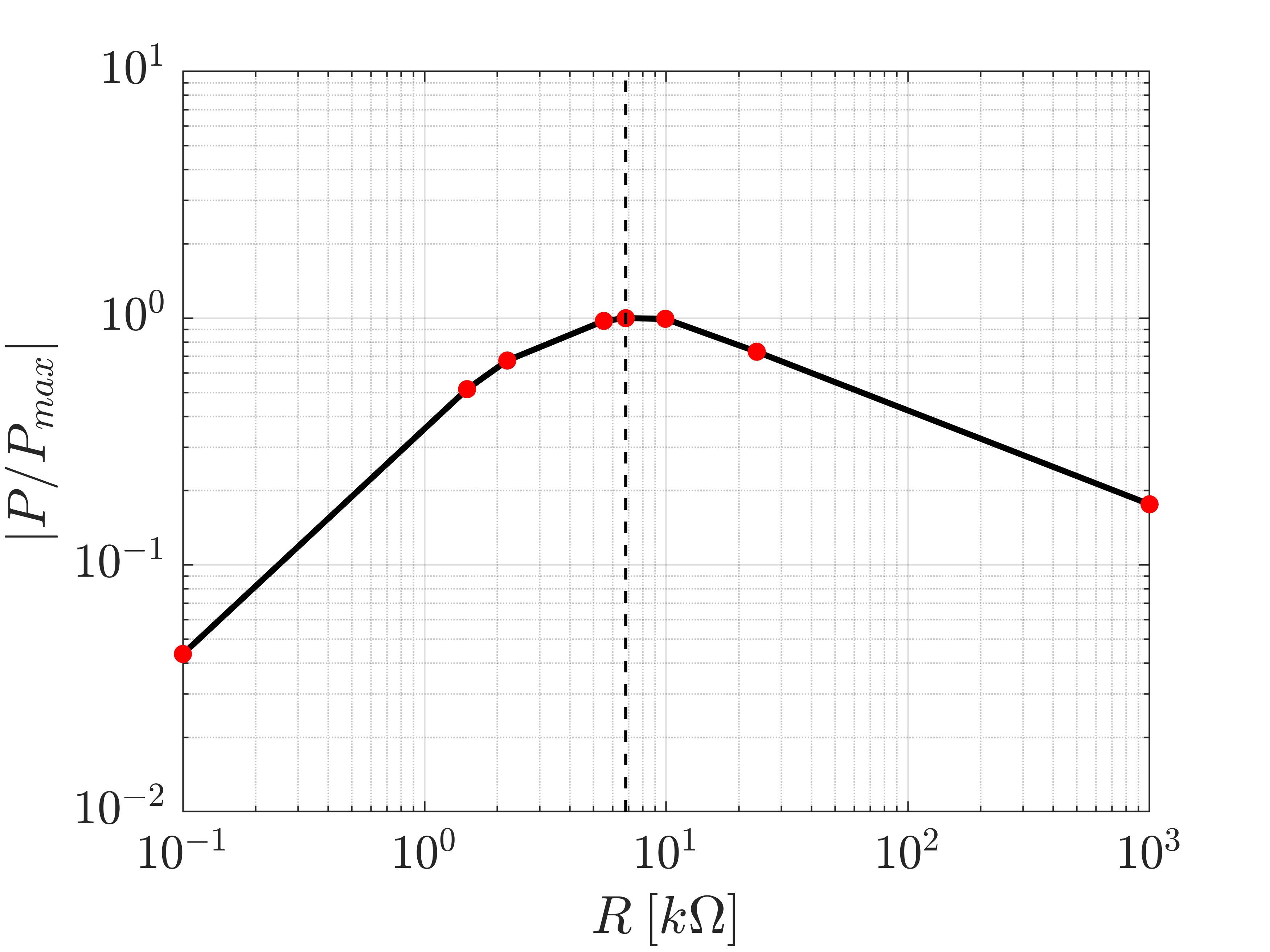} \label{FigS1a}}	\caption{(a) Experimental setup adopted for the evaluation of the optimal electric load and associated electric circuit  (I). The same load and electric connections are employed in the experimental tests of the beam with attached resonating rod (II) and beam with metasurface. Zoom-in view of the harvester (III). (b) Representation of the peak power for different resistive loads. The optimum values is approximately $6.8$ $k\Omega$.}
	\label{Fig:OptimalLoad}
\end{figure*}

\noindent
\textit{Geometrical and physical details of the harvester and optimal resistive load}\\
\\
The piezoelectric harvester employed in this work is made of four cantilever beams mounted on the $26^{th}$ resonator. Each of them is made of an aluminum substrate of $25$ mm length, $5$ mm width and  $2$ mm thickness, endowed with piezoelectric patch of the same in plane dimensions, with a cross-section thickness of {$t=0.3$ mm} and bonded through a conductive epoxy CW2400. 
To optimize the power extracted during the tests, a passive resistor is selected and connected to each piezo-patch based on the experimental characterization of the dynamical response. 
The analysis is performed through the experimental setup displayed in Fig. \ref{Fig:OptimalLoad}(a)-I, where the resonator is placed vertically on the shaker. The system is excited using a sweep sine in the neighborhood of the resonance frequency of the cantilever and the rod (which are carefully designed for operating at the same frequency). The voltage drop across the piezoelectric electrodes is measured upon varying the resistance values, whereby the power achieved at $2.05$ kHz is displayed in Fig. \ref{Fig:OptimalLoad} (b) normalized by the its maximum. The analysis reveals an optimal resistance of $6.8$ $k\Omega$, which is in agreement with the lumped model in Ref. \cite{DePonti}. 

\section{SUPPLEMENTARY NOTE 2 - NUMERICAL METHODS AND ADDITIONAL RESULTS}
The numerical model employed in this work is based on a finite element discretization of the system through ABAQUS CAE 2018$^{\textregistered}$. In particular, the host beam and the resonators are modeled through full 3D stress quadratic elements (C3D20), while the electromechanical interactions are addressed by linear 3D piezoelectric elements (C3D8E). Electrodes are modelled through a voltage constraint on the piezo-faces, and the electric load is introduced via a user Fortran subroutine, that is able to solve at each time increment a resistive circuit \cite{Gafforelli1}.\ The analysis is performed using an implicit analysis based on the Hilber-Hughes-Taylor operator, with a constant time increment $dt = 5$ $\mu s$.
The system is forced using the imposed acceleration that is experimentally measured with the accelerometer placed in correspondence of the shaker, while the black hole is modeled numerically by varying the beam profile and adding a soft material with Rayleigh damping $\alpha = 5000$ rad/s. In addition, we add in the harvester model a Rayleigh damping source, with $\alpha=19.3$ rad$/$s and $\beta=2.93\times 10^{-6}$ s$/$rad on the basis of conventional modal damping ratios \cite{ErturkBook} for aluminum beams with piezoelectric patches.\\ 
The dispersion relation is computed using COMSOL multiphysics$^{\textregistered}$ exploiting periodic boundary conditions applied to the unit cell boundaries, whereby a parametric analysis is performed in order to verify the wavenumber transformation described in Fig. 2(b) and Fig. 4(d) in the manuscript.

\begin{figure*}[h!]
	\centering
	\hspace{-0.4cm}
	\subfigure[]{\includegraphics[width=0.45\textwidth] {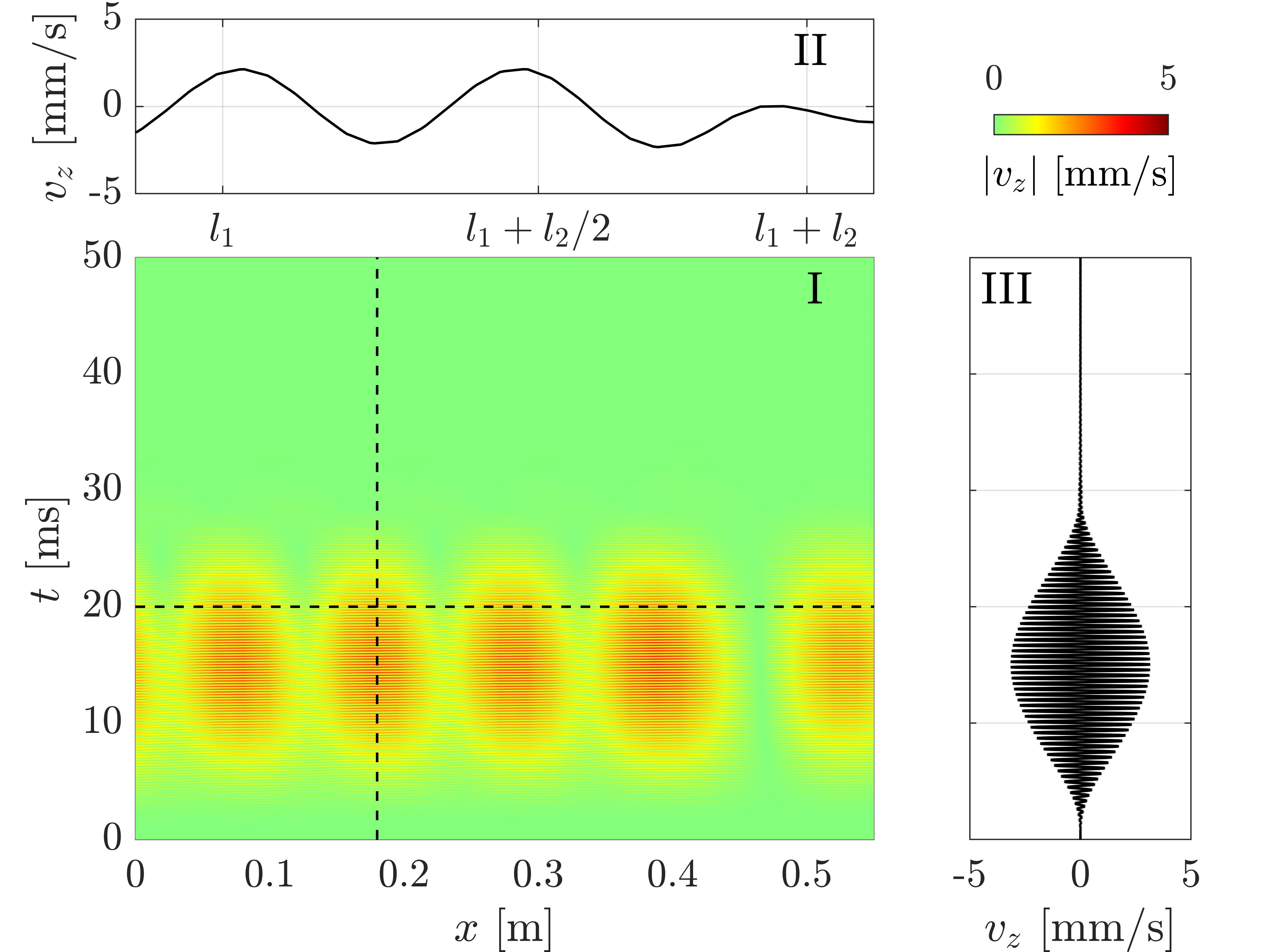} \label{FigS1a}}
	\subfigure[]{\includegraphics[width=0.45\textwidth] {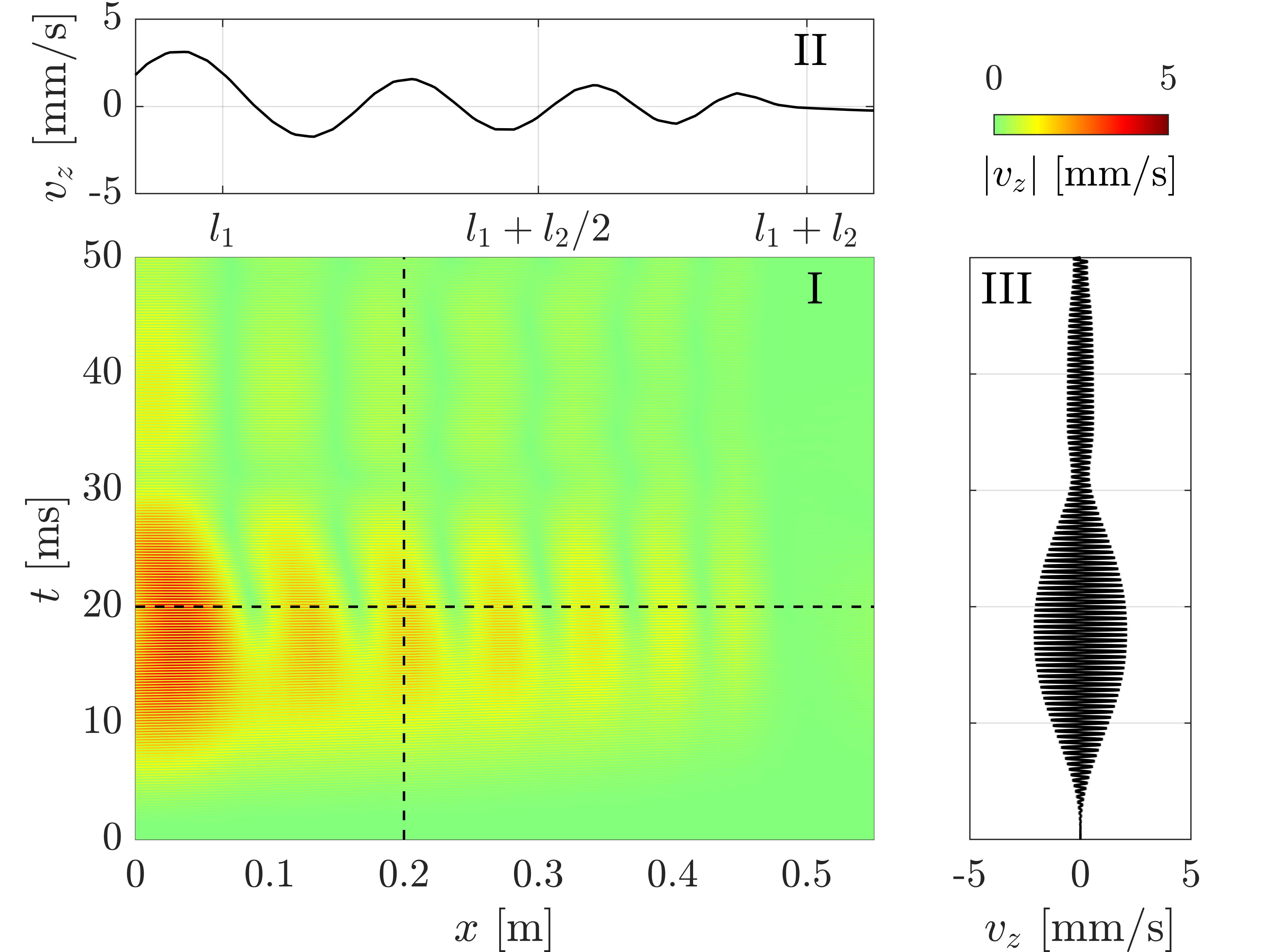} \label{FigS1a}}	\caption{Numerical velocity field computed for (a) beam with one resonator; (b) beam with graded metasurface. The horizontal and vertical dashed lines in (${\rm I}$) correspond to the reference coordinate and time instant employed to illustrate the wave behavior in space (${\rm II}$) and time (${\rm III}$), as shown on top and alongside the velocity field. }
	\label{FigS1}
\end{figure*}